\documentclass[12pt,preprint]{aastex}
\def\gapprox{\;\rlap{\lower 2.5pt            
    \hbox{$\sim$}}\raise 1.5pt\hbox{$>$}\;}       
\def\lapprox{\;\rlap{\lower 2.5pt            
    \hbox{$\sim$}}\raise 1.5pt\hbox{$<$}\;} 
\begin{document}

\title{Energy Distribution of Micro-events in the Quiet Solar Corona}

\author{Arnold O. Benz}
\affil{Institute of Astronomy, ETH-Zentrum,
    CH-8092 Z\"urich, Switzerland}

\and

\author{S\"am Krucker}
\affil{Space Sciences Laboratory, University of California, Berkeley CA 94720, USA}

\begin{abstract}
Recent imaging observations of EUV line emissions have shown evidence for frequent flare-like events in a majority of the pixels in quiet regions of the solar corona. The changes in coronal emission measure indicate impulsive heating of new material to coronal temperatures. These heating or evaporation events are candidate signatures of "nanoflares" or "microflares" proposed to interpret the high temperature and the very existence of the corona. The energy distribution of these micro-events reported in the literature differ widely, and so do the estimates of their total energy input into the corona. Here we analyze the assumptions of the different methods, compare them by using the same data set and discuss their results. 

We also estimate the different forms of energy input and output, keeping in mind that the observed brightenings are most likely secondary phenomena. A rough estimate of the energy input observed by EIT on the SoHO satellite is of the order of 10\% of the total radiative output in the same region. It is considerably smaller for the two reported TRACE observations. The discrepancy can be explained partially by different thresholds for flare detection. There is agreement on the slope and the absolute value of the distribution if the same method were used and a numerical error corrected. The extrapolation of the power law to unobserved energies that are many orders of magnitude smaller remains questionable. Nevertheless, these micro-events and unresolved smaller events are currently the best source of information on the heating process of the corona.
\end{abstract}


\section{Introduction}
The heating of the solar corona has been a riddle since the discovery of the high coronal temperature by Edl\'en and Grotrian in the late 1930's. Does the heating process leave any other signature? There are many proposals on possible sources of energy (e.g. reviews in Ulmschneider, Rosner \& Priest 1991), but proof can only come from observations of the corona. Of particular interest are temporal variations in coronal temperature, density and energy content, as they may reveal clues on the nature of any non-stationary heating process.

In the quiet corona on which we concentrate here, small brightenings above the network of the magnetic field were first discovered in deep soft X-ray exposures by Yohkoh/SXT (Krucker et al. 1997). The events were reported to have a typical thermal energy of $10^{26}$erg, and to occur at a rate of 1200 per hour extrapolated over the whole Sun. These micro-events seem to be present at all times (Pre\'s et al. 2001) and also in coronal holes (Koutchmy et al. 1997). Thus they represent a population of continuous dynamic coronal phenomena, quite different from the more sporadic regular flares occurring in active regions that are strongly related to solar activity and its cycle. Micro-events also appear different from X-ray bright points discovered by Skylab that emerge at a rate of 62 per hour averaged over the whole Sun and persist on average for 8 hours (Golub et al. 1974). The radiative energy loss during the bright point's lifetime is of the order of $10^{28}$ erg. They seem to be the place of frequent micro-events (Pre\'s \& Phillips 1999). More sensitive measurements of micro-events became possible with EIT and TRACE. Benz \& Krucker (1998) and Berghmans, Clette \& Moses (1998) found independently that the coronal emission measure in quiet regions observed in EUV iron lines fluctuates locally at time scales of a few minutes in a majority of pixels including even the intracell regions. At the level of 3 standard deviations, Krucker \& Benz (1998, thereafter KB) reported the equivalent of 1.1$\times 10^6$ coronal micro-events per hour over the whole Sun for SoHO/EIT observations. Benz \& Krucker (1998) noted a nearly linear relation between inferred averaged input power and radiative loss per pixel.

In the transition region of the quiet sun, line emissions have been observed to have localized brightenings and explosive events (Brueckner \& Bartoe 1983, Dere 1994) at a rate increasing with instrumental sensitivity (Porter et al.1987; Harrison 1997). Winebarger et al. (2001) estimate that explosive events observed in SoHO/SUMER line broadenings contain an upward energy flux corresponding to about 10\% of the required energy to heat the corona. Habbal (1992) and Brkovi\'c et al. (2000) have pointed out that some of them are associated with a coronal brightening, others are not. Here we concentrate on events that have an impact on the corona. Thus we search for an observable change of the energy density in localized regions in the corona. 

The energy of coronal micro-events can be estimated from their peak temperature and emission measure or from enhanced line intensities integrated over events. The two methods are orthogonally different, the first measuring the input, the second the output. The two methods split again by the choice of the spatial association: {\sl (i)} The enhancements in individual pixels, thus in a given area, has been previously analyzed. The density and thermal energy of the new material are difficult to estimate, as the depth of the event cannot be observed. {\sl (ii)} The energy inputs into adjacent brightening pixels can be added to events. The combination of active pixels to events requires some non-trivial assumptions. The advantage of integrating over events lies in the possibility to estimate the depth from shape and size of the projected area. These choices constitute 4 completely different ways to derive the energy input by micro-events. The resulting distributions are critically reviewed and compared below.

The terms "microflare" and "nanoflare" have been introduced as theoretical concepts referring to subresolution reconnection events (Parker 1983). Soon after, observers used them for small flare-like brightenings below previous thresholds in active regions (e.g. Lin et al. 1984; Gary et al. 1997). Here we use the term "micro-event" to denote short enhancements of coronal emissions in the energy range of about $10^{24} - 10^{27}$erg in quiet regions. The lower limit is given by current instrumental thresholds, and the upper limit refers to the largest events occurring typically in a quiet region of a few arcminutes size within one hour. 

In Section 2 we discuss the variability of the quiet corona and present examples of individual pixels for illustration; the various energies involved are discussed and compared. Event selection and energy distributions are assessed in Section 3. We use the data originally presented by KB and Benz \& Krucker (1998) as the base to critically review the published event definitions and to investigate their effects. The results are discussed in Section 4. Section 5 gives conclusions and suggestions for future work.

\section{The Observational Basis}
Temporal changes of coronal emissions are ideal for testing heating models. In this section, observations of such changes are presented. The energy associated with events is often inferred from the emission measure, ${\cal M}$, of soft X-rays or coronal EUV lines. We define ${\cal M}:=\int n^2 dV\approx \int n^2A\ ds$, where the integration in $s$ is along the line of sight, $A$ is the area or pixel size, and the density $n$ refers to the plasma in the temperature range given by the observed emission process. The emission measure is proportional to the emission in a line for a given temperature. 

In the following, the emphasis is on emission lines observed in the two wavelength bands, 171 \AA \ and 195 \AA, including lines of Fe IX/X and Fe XII, respectively, with diagnostic capabilities in the range of plasma temperatures of about 1.1-1.9$\times 10^6$K. These coronal lines dominate the observed parts of the EUV spectrum, and their large photon fluxes provide higher sensitivity than previous observations in soft X-rays. Using lines of different ionization states, the emission measure in a range of temperatures can be derived. The ratio of the emission in two lines also defines an isothermal temperature. 

The derived parameters are to be taken as formal values, representing weighted means over the sensitive temperature range. The line-ratio temperature and emission measure have been determined for each pixel at each time step from observations by the Extreme ultraviolet Imaging Telescope (EIT) on SoHO (Delaboudini\`ere et al. 1995) and TRACE (Handy et al. 1999).

\begin{figure}
\plottwo{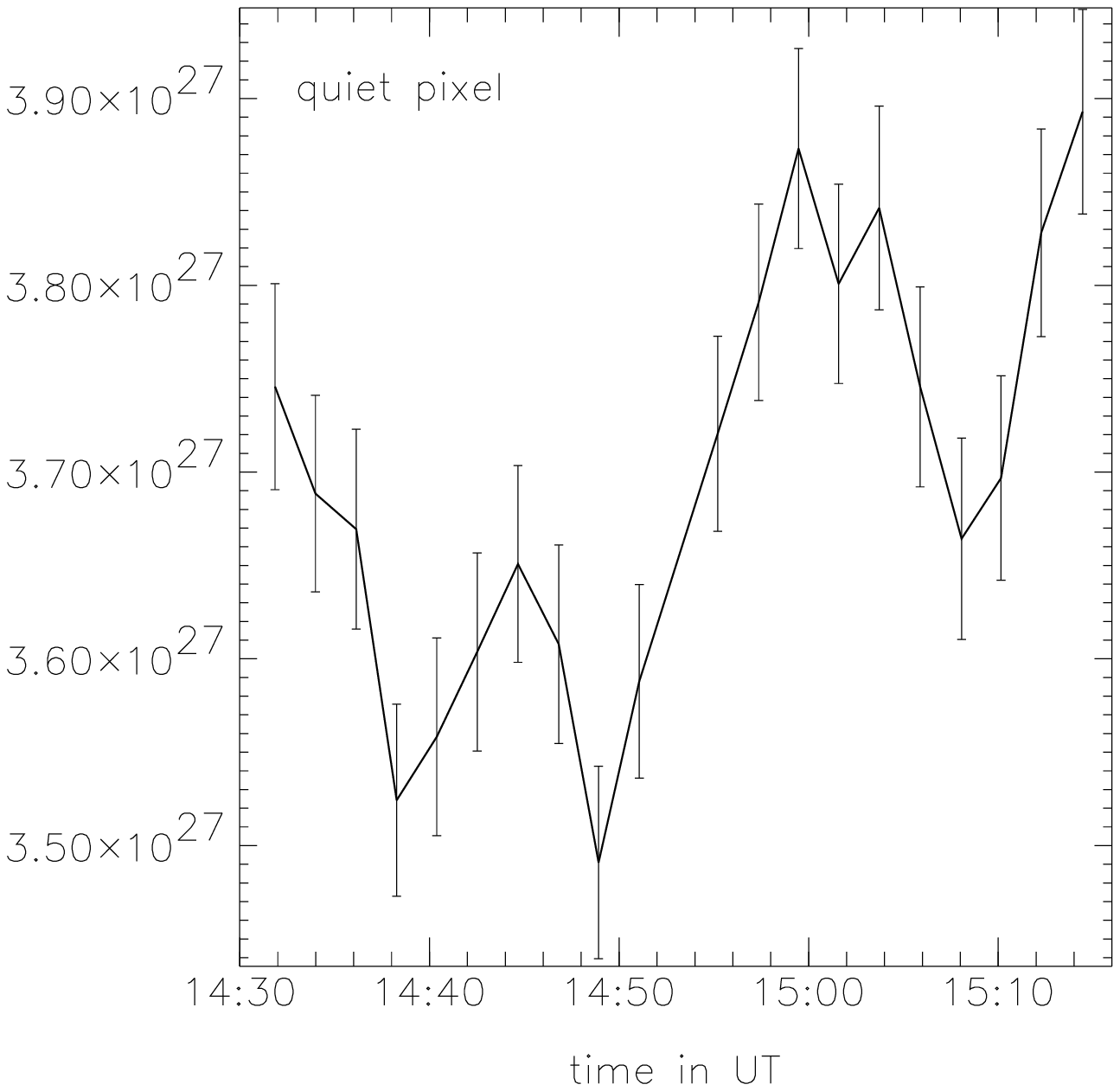}{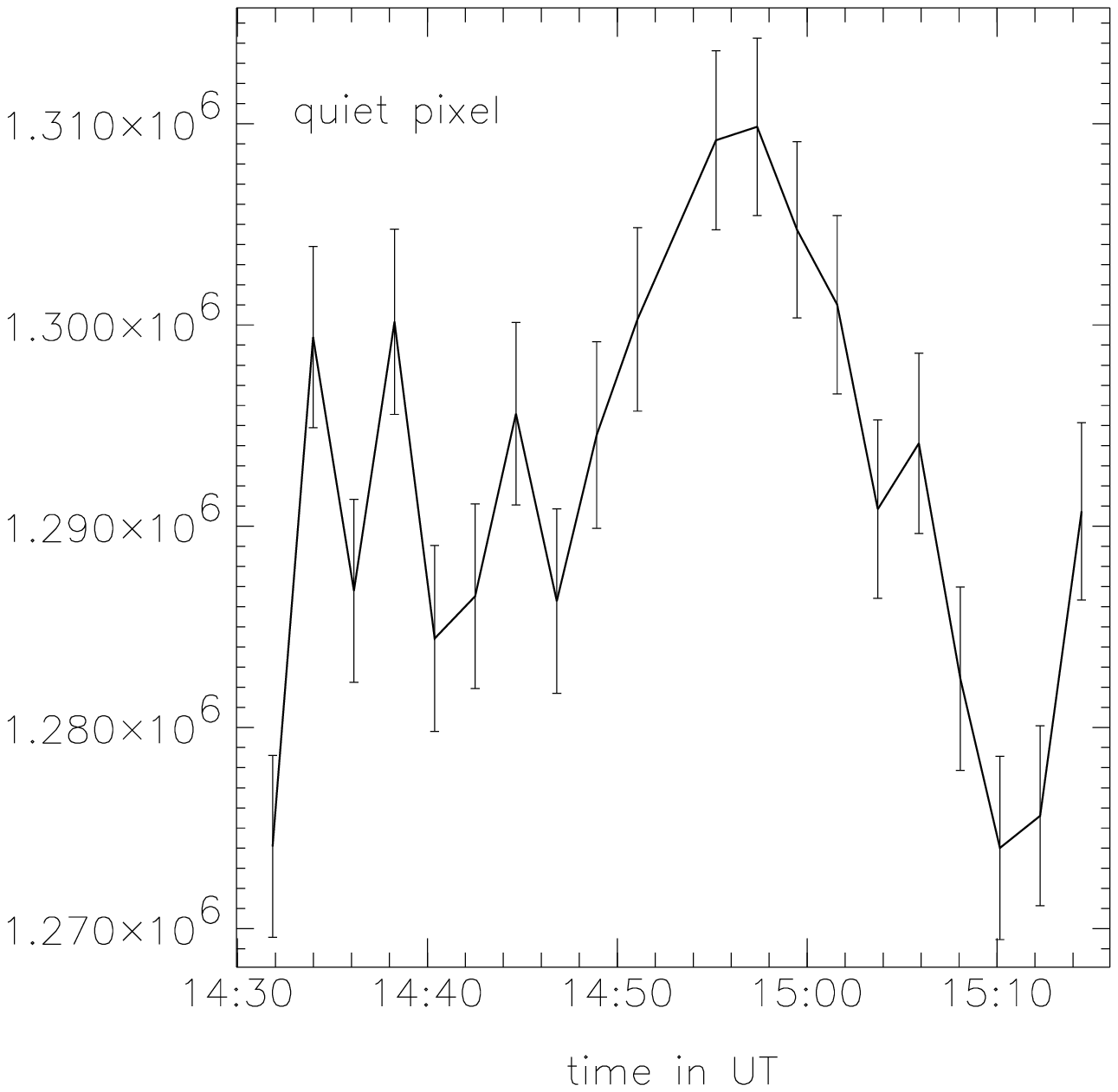}
\caption[f1a.eps]{{\sl Left:} Coronal mission measure of a random 1900km$\times$1900km pixel in a quiet region of the Sun as determined from EIT high-temperature iron lines. The emission measure is divided by the pixel area $A_p$. The error bars are conservative upper limits as explained in the text. {\sl Right:} The formal temperature evolution of the same pixel is shown. No background is subtracted. \label{fig1}}
\end{figure}

\subsection{Observed Variability}
The variability of the quiet corona near the center of the disk is illustrated in Fig. 1. The presented observing run lasted from 14:30 to 15:15 UT on July 12, 1996, when solar activity was almost at the lowest level during the most recent solar minimum. The time resolution is 127.8 s and the pixel size 2.62" (1900 km on the Sun). The error bars of the emission measure have been estimated from the low $\sigma$ values of distribution of the largest enhancements in the time series (Fig.2). As there are non-gaussian variations clearly visible in Fig.2, these error bars are upper limits. They are also the basis to estimate the accuracy of the temperature (Fig.1, right).

The pixel shown in Fig. 1 is typical for a location above the magnetic network. The emission measure increases significantly twice within 42 minutes with peaks at 15:00 and $\gapprox$ 15:14 UT. A third event at 14:45 UT is marginal. In the event at 15:00 UT the temperature peaks shortly before the emission measure in agreement with previous reports (e.g. cross-correlation by Benz \& Krucker 1999). 

\begin{figure}
\plotone{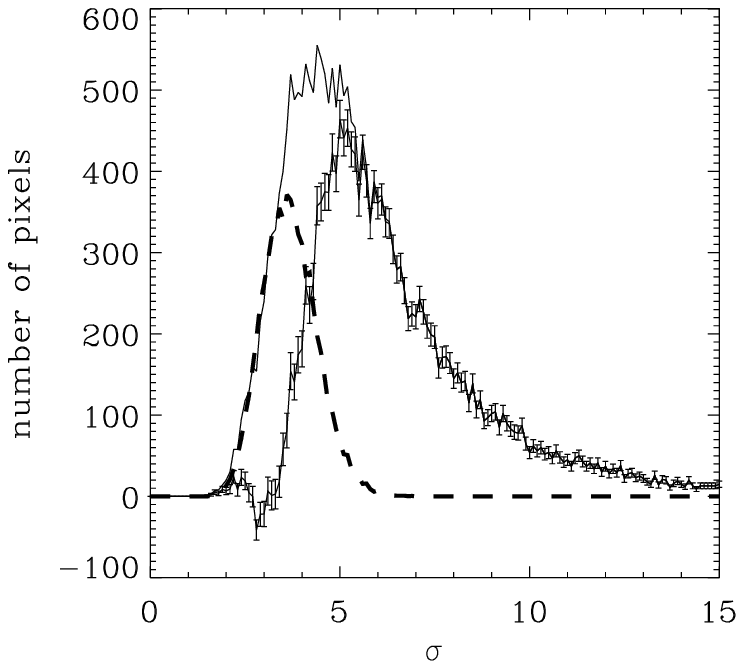}
\caption[f2.eps]{Distribution of pixels (thin curve) having emission measure enhancements (maximum - minimum) larger than the indicated $\sigma$-value in the 42 minutes observing time. EIT observations of 23800 pixels in a 7'$\times$7' quiet area are presented. The dashed curve represents the distribution expected from gaussian noise and was fitted at the low-$\sigma$ values (0 - 3 $\sigma$). The difference between the observed distribution and the noise is the curve shown with error bars. \label{fig2}}
\end{figure}

Statistics on the variability of all pixels are presented in Fig. 2. The values are normalized to the standard deviation of each pixel. It was derived in the following way (Benz \& Krucker 1998): First, the average photon noise $\sigma_{\rm phot}$ in each pixel was evaluated and the distribution was plotted in these units. Then a distribution calculated from gaussian noise was fitted to the 0 - 3 $\sigma_{\rm phot}$ part. The noise it represents is the maximum limited by the condition that it cannot be larger than the observed distribution plus 3 times the standard error. The fitted curve represents an effective $\sigma$, 20\% larger than $\sigma_{\rm phot}$. Finally Fig. 2 was plotted in units of $\sigma$. 

There is a non-gaussian, enhanced tail in the distribution of the difference between maximum and minimum during the time series (Fig.2). As the fit may include small noise-like, but real fluctuations, the fitting curve is an upper limit of the noise contribution. The fraction of the pixels with non-gaussian deviations is 72\% (ratio of integrals under dashed curve to full curve in Fig.2). This percentage is a lower limit on the pixels with significant changes of the emission measure. 

The surface of the network covers about 10\% of the quiet sun (Rutten \& Hogenaar 2001). Since more than 72\% of the pixels in a quiet region brighten within 42 minutes, the statistics in Fig.2 indicates that many intracell regions vary significantly over one hour, confirming previous results (cf. Sect. 1). 

\begin{figure}
\plotone{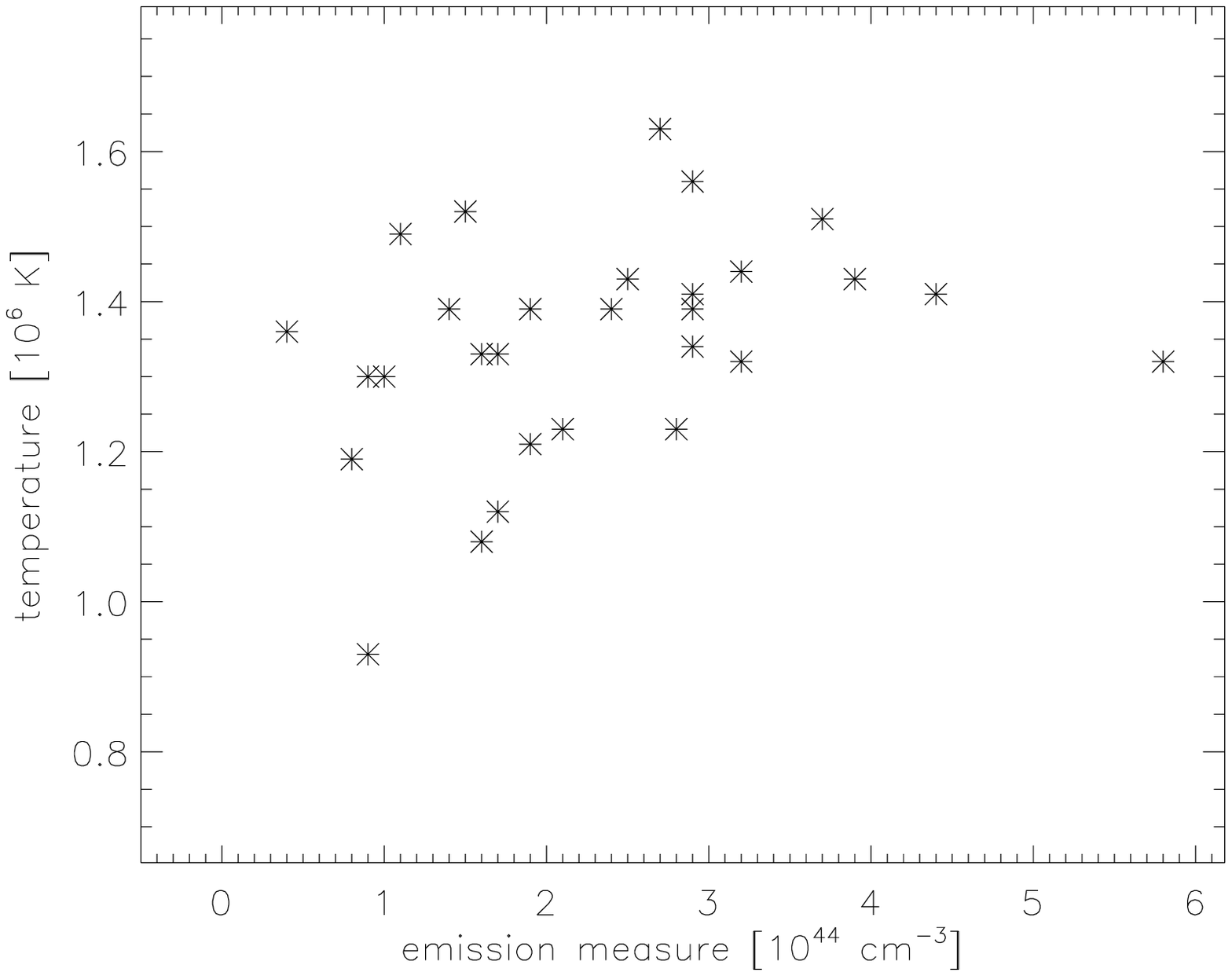}
\caption[f3.eps]{The increase of the emission measure vs. average line-ratio temperatures of relatively large micro-events in the quiet corona determined from the line ratio of Fe IX to Fe XII observed by EIT. The pre-event background flux in each line was subtracted before the temperature of the integrated events was evaluated. \label{fig3}}
\end{figure}

The temperature of the material causing the emission measure increase is best derived from the largest events. Figure 3 shows the temperature in the 29 largest events that occurred in a 7'x7' field within 42 minutes (data from Krucker \& Benz 2000). The observed temperature range is from 0.9 to 1.6$\times 10^6$K. The lower limit may originate from the temperature sensitivity of the lines used.  The upper limit of the observed range is not limited by the observing method. It is consistent with the temperatures derived from Yohkoh/SXT using different filtergrams of the soft X-ray continuum (Krucker et al. 1997). The average temperature of the events plotted in Fig.3 is 1.34$(\pm 0.15)\times 10^6$K at an average emission measure of 2.3$\times 10^{44}$cm$^{-3}$.

A slight and apparently significant correlation between temperature and emission measure is discernible in the micro-events of Fig.3 (cross-correlation coefficient $r$ = 0.34, $n$ = 29, significance $p > 95$\%). However, the slope of the regression line is relatively flat: 
\begin{equation}
T\ =\ 1.246 + 0.041 {\cal M}_{44} \ \ {\rm [}10^6{\rm K]}\ \ .
\end{equation}
The determination of the above correlation significance does not include the errors of individual measurements. They are of the order of 5\% for the emission measure and 10\% for the temperature. If they are included, the data displayed in Fig. 3 are compatible with no relation between emission measure and temperature.

\subsection{Energy Estimate of Events}

\noindent{\sl Radiation Energy}\par
The energy of micro-events has been estimated from the radiation output in EUV lines. The emission in one line has to be integrated over the event and the total radiation of the plasma (say 1 - 300\AA ) to be inferred from some temperature estimate and an emissivity code (such as SPEX or CHIANTI). Note that the radiation is considered only during the time the plasma is within the range of sensitive temperatures. Other losses, in particular conduction, may also have to be taken into account to estimate the total energy output. Thus the radiative output underestimates the energy loss. Measuring the radiation output has the advantage that the line of sight thickness of the micro-event does not have to be modeled. The disadvantage is the assumption of the temperature.

\noindent{\sl Thermal Energy}\par
A completely different way to estimate the energy measures the thermal energy content at a given time. One observes two lines and determines the temperature from the line ratio. Using the same element at two ionization states, the line ratio defines an average temperature $T$. This allows the determination of an emission measure relevant for the plasma within the range of temperatures appropriate for the two lines. As the measurements are made at peak flux, more events in a larger energy range can be studied. The results are more propitious because the information from two lines is used. 

The thermal energy of the material exceeding a given threshold $T_0$ is
\begin{equation}
E_{\rm th} \approx 3 k_B T \int {n_e(T>T_0)\ dV}\ \ \ \ .
\end{equation}
Eq. (2) includes the contributions of both electrons and ions, assumed to have the same temperature. The electron density in the source region, $n_e\approx n_i$, is determined from the observable increase in emission measure $\Delta {\cal M}$,
\begin{equation}
\Delta {\cal M} = \int {[(n_{e,0} + n_{e,1})^2 - n_{e,0}^2] dV}\ \ \ ,
\end{equation}
where $n_{e,0}$ is the background density before the event and $n_{e,1}$ is the density of the newly added material. For simplicity let us consider here constant densities in the volume $V_0$ before the event and in $V_1$, the volume into which the new material is injected. The emission measure can easily be evaluated for two extreme cases yielding the same observed enhancement: {\sl (i):} $ n_{e,1}\ll n_{e,0}$ in the same volume,  $V_0 = V_1$, i.e. low density material is newly injected into the old volume, and {\sl (ii):} $ n_{e,1}\gg n_{e,0}$, i.e. high density material is injected into a small part of the old volume, $V_1 \ll V_0$. In the first case the density of the newly added material can be estimated from

\begin{equation}
n_{e,1} \approx {{\Delta {\cal M}}\over {2\sqrt{{\cal M}_0 V_0}}}\ \ \ .
\end{equation}
In the second case the density of the newly added material is

\begin{equation}
n_{e,1} \approx \sqrt{\Delta {\cal M}\over V_1}\ \ \ .
\end{equation}
Brown et al.(2000) have evaluated some realistic models in detail and have found differences in the derived densities by a factor of two. We will assume the second case in the following and write $V_1 = qV_0 = A s_{\rm eff}$, where $q$ is the filling factor of the newly added material and $A$ is the observed area. The parameter $s_{\rm eff}$ represents an effective line-of-sight thickness of the new material. It includes the filling factor for multi-pixel events and corrects also for cases with smaller event area than the pixel size. It is a model parameter and has been assumed to be 5000 km in KB, which is probably an overestimate for small events. A lower limit may be estimated from the assumption that the smallest observable events extend about the length of a pixel (2000 km) with half its diameter (1000 km), thus $s_{\rm eff} \gapprox $ 500 km. 

Combining Eqs. (2) and (5), the change in thermal energy amounts to 
\begin{equation}
E_{\rm th} \approx 3 k_B \Delta T\sqrt {\Delta {\cal M} A s_{\rm eff}}\ \ \ .
\end{equation}
Enhancements both in emission measure and temperature occur in Fig.1. The thermal energy of the corona can increase either by adding material previously "invisible" (e.g. heated chromospheric material) or by increasing the temperature of coronal material. For illustration we evaluate the event peaking at 15:00 UT. The increase $\Delta{\cal M} = 3.5\times 10^{26} A_p\ $cm$^{-3} $ indicates heating of chromospheric material, thus $\Delta T\approx T$. Even for an $s_{\rm eff}$ of only 500 km and without subtracting a background, the required energy is $2.6\times 10^{24}$erg. Thus the example suggests that heating new material to coronal temperatures in micro-events is a major energy input. Other temperature variations in Fig.1 are of the order of $10^4$K. The average emission measure of the pixel in Fig.1 is $3.7\times 10^{27} A_p$ cm$^{-3}$ where $A_p$ is the pixel area. At a constant emission measure the temperature variations would amount to less than 10\% of the above energy. Temperature variations at constant emission measure thus constitute negligible heating. Surprisingly, the dominant variability of the coronal energy content is not in temperature but in emission measure.

The reported emission measure increases have been measured at peak value, but conduction and radiation losses occur also during the rise phase of an event. As the rise time in micro-events is comparable to the decay time, about half of the radiative loss occurs before the peak. Thus the thermal energy at peak underestimates the total energy input by at least a factor of two. More detailed calculations on this problem have been presented by Brown et al. (2000). 

\noindent{\sl Gravitational Energy}\par
The gravitational energy of relatively large heating events has been estimated by several authors (e.g. Brown et al. 2000). It amounts to a few percent of the thermal energy estimated above and will be neglected in the following. 

\noindent{\sl Isothermal Expansion}\par
For an expansion along a flux tube of constant diameter $a$, the electron density decreases inversely proportional to the enlargement. We assume a homogeneous volume with the initial density $n_{e,0}$ in the chromosphere expanding homogeneously to the coronal value of $n_{e,1}$, where  $n_{e,0} \gg n_{e,1}$ and $ n_{e,1} = n_{e,0} \ell_0/\ell_1$. The temperature is not seen to decrease from a very high value, thus the expansion may be modeled as being isothermal. The expansion energy can then be approximated to

\begin{eqnarray}
E_{\rm exp}\ \ =& \int p\ a\ d\ell\ \  &\approx \ \ 2 k_B T_1 a\int n_e  d\ell \\
=& 2 k_B T_1 n_{e,1} a \ell_1 &\ln{(\ell_1/\ell_0)}\ \ \ \approx\ \ 2 \ln{(\ell_1/\ell_0)}k_B T_1\sqrt {\Delta {\cal M} A s_{\rm eff}}\ \ \ ,
\end{eqnarray}
where $p$ is the total particle pressure, and $\ell_0$ and $\ell_1$ are the initial and final extensions, respectively, of the newly heated material along the loop. The expansion energy exceeds the thermal energy by the factor $\gamma \approx 2/3 \ln(\ell_1/\ell_0) > 1$. For realistic values of chromospheric and coronal densities, this factor is about 3. The energy is liberated when the injected material cools and concentrates again in a smaller volume. As most of the micro-events in the quiet corona seem to take place in closed loops, this is the case for most of the expansion energy.

The combined input power including heating and expansion may be written in terms of $P_{\rm th} = \Delta E_{\rm th}/ \Delta t$, where $\Delta E_{\rm th}$ is the observed input of thermal energy in the observing time $\Delta t$
\begin{equation}
P_{\rm inp}\ \ \approx\ \ \alpha (\beta P_{\rm th}\ +\ \gamma P_{\rm th}) \ =\ P_{\rm out}\ \ \approx\ \ \mu P_{\rm rad}^{>1MK}.
\end{equation}
The factor $\beta \approx 2$ covers the losses before the emission measure is determined at its peak. It may be noted here that the sum $\beta+\gamma\approx5$ is reduced in case of a small filling factor of the injected material. The factor $\alpha$ takes into account that the observations are based on the secondary effect of heating and evaporation. As primary plasma motions and waves may result from the energy release, $\alpha \geq 1$. For reconnection, $\alpha \approx 2$ has been estimated in the MHD limit (e.g. Priest \& Forbes 2000).

Eq.(9) equates finally the input to the output. In the previous literature, the radiative output $P_{\rm rad}$ has been estimated from the observed coronal emission measure ($\gapprox 10^6$K) and the line-ratio temperature. This ignores the radiation at lower temperature and underestimates the total radiative loss by some factor $\mu>1$.  If the coronal material cools from 1.5$\times 10^6$K to chromospheric values, $\mu \approx 3$.

\begin{table}
\begin{center} 
\caption{Reported power-law indices of the energy distribution of micro-flares. Literature references to Krucker \& Benz; Berghmans, Clette \& Moses; Parnell \& Jupp; Aschwanden et al.; Pre\'s, Phillips \& Falewicz. Power-law index $\delta$, method for energy determination (em = emission measure increase; rl = radiation loss), time window in minutes for combining pixels to events (synchrony), height model for density determination, instrument, observed coronal emission line, flare selection.} 
\begin{tabular}{llllllll} \hline
& & & & & & & \\
reference & $\delta$& method &synchr.& height & instrument&emission &selec-\\
  &  & & [min] &model & &line &tion\\
& & & & & & &\\
\hline
KB (1998)&2.59 &em&2 &h= const.&EIT &Fe IX/XII&none\\
BCM (1998)&1.35 &rl&event&none &EIT &Fe XII &none\\
PJ (2000)&2.52&em&2&h= const.&TRACE&Fe IX/XII&none \\
PJ (2000)&2.04&em& 2&h=$\sqrt{A}$ &TRACE &Fe IX/XII&none \\
AEA (2000)&1.80&em& 6&h=$\sqrt{A}$  &TRACE &Fe IX/XII&selection  \\
PPF (2001)&2.47&em& ?&?  &Yohkoh &continuum&none  \\
This work&2.31 &em&2 &h=$\sqrt{A}$  &EIT &Fe IX/XII&none \\
& & & & & & & \\
\hline
\end{tabular}                                             

\end{center}
\end{table}

\section{Event Selection and Energy Distributions}

\noindent{\sl Radiative Output}\par
Table 1 summarizes the published results on the power-law index of energy distribution. The energy distribution derived from radiation loss by Berghmans et al. (1998) is remarkably flat compared to the others. The global event rate reported by these authors is 4300 per hour, compared to 1.1$\times 10^6$ by KB. In both studies SoHO/EIT data were used. Berghmans et a. (1998) took a local 3$\sigma$ peak and combined to one event all spatially adjacent pixels that brightened more than 2$\sigma$. Each of these pixels was considered to be part of this event as long as it stayed above 2$\sigma$ during the brightening of the first pixel. The events defined in this manner reach five times larger areas than in KB and possibly include small independent events. Thus the difference mainly originates in how brightened pixels are combined to events rather than in the different way the energy is measured. 

\noindent{\sl Thermal Input}\par
In the following we concentrate on energy estimates from thermal input and discuss the errors concerning the energy distribution. These are relative errors that arise from comparing energies of different events. It is clear from Section 2.2 that the estimate of the total input of an event is still rather uncertain in absolute terms. 

\subsection{Event Selection}
For event selection a simple procedure based on a lower limit given in terms of $\sigma$ is applied in most cases (e.g. KB; Parnell \& Jupp 2000). A local maximum is searched in the timeline of a pixel. The preceding local minimum or the start of the series is taken as the background level. 

Aschwanden et al.(2000a) have used more restrictive selection criteria. For each pixel they determined the absolute peak in the time series and took the absolute minimum as the background. Thus they accepted only one event per observing time and per pixel, excluding cospatial events. This selection is biased against small events. Secondly, they required the temporal cross-correlation coefficient between both lines be larger than 0.5 to exclude non-flare-like events. As small events have a smaller signal-to-noise ratio and thus are more likely to be excluded, the criterion again favors large events. The second criterion excludes more than 70\% of the detected events.

That an event selection as defined by Aschwanden et al.(2000a) is problematic, can be seen in their two completely different event distributions depending on the wavelength in which they select events. The cross-correlation criterion is intended to select only events of the same type, seen clearly at both wavelengths. Hence the selection should give similar flare distributions. This is not the case as their Fig. 12 demonstrates.

Are there several kinds of coronal micro-events, flare-like and non-flare-like? The temperature distribution in Fig. 3 does not suggest different populations of events in EIT data. The energy distributions reported by the various groups decrease continuously with energy; no secondary peaks have ever been noted. The temporal cross-correlation of Fe IX and Fe XII averaged over all pixels is excellent ($r$ = 0.46 at an Fe IX delay of 23 seconds, Benz \& Krucker 1999), indicating that the variations in the two lines are well correlated except for noise. We do not find a reason to distinguish between events with different line ratios, i.e. different temperatures. 

Both additional selection criteria introduced by Aschwanden et al. (2000a) exclude predominantly small events and therefore reduce the power-law index $\delta$. The second criterion alone flattens the distribution by 0.3 in the power-law index (Aschwanden et al. 2000b).

\begin{figure}
\plotone{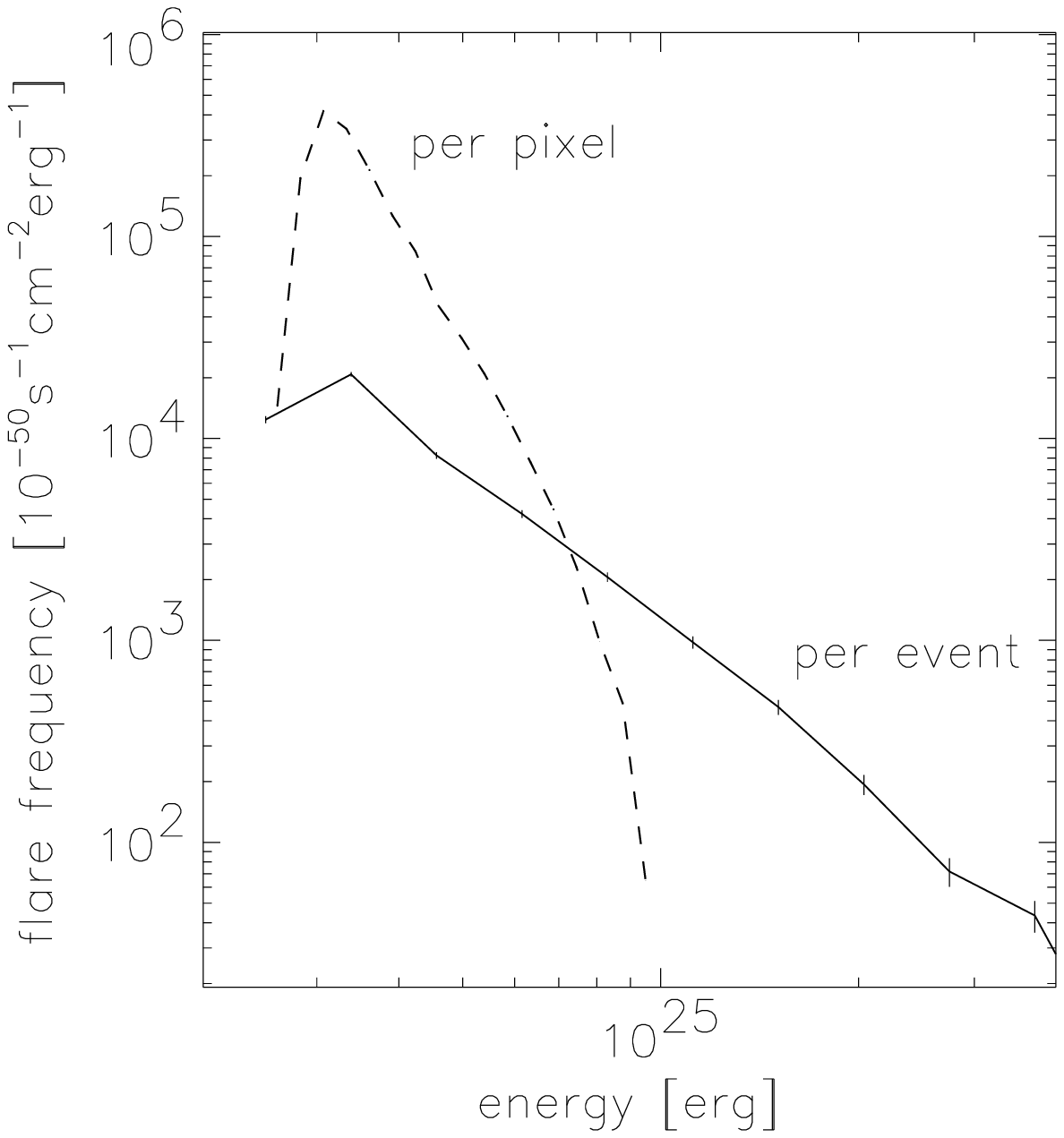}
\caption[f4.eps]{The energy distribution of emission measure enhancements in individual pixels (height model: $s_{eff}$=500km; dashed) observed in the quiet corona is compared with the distribution of micro-events (full curve with error bars), in which all adjacent pixels brightening simultaneously have been combined (tolerance $\pm$1 minute; $s_{eff}=\sqrt{A}$). \label{fig4}}
\end{figure}

\subsection{Combining Pixels to Events}

Figure 4 shows the energy distribution of emission measure enhancements in single pixels. An effective line-of-sight depth of 500 km has been assumed (cf. Eq.6). The power-law index in the middle part is $\delta$ = 6.04. At high energies, the distribution falls off with a maximum energy at 10$^{25}$erg. The low-energy cut-off is instrumental. Combining adjacent pixels to events flattens the distribution and extends the maximum energy to 4$\times 10^{26}$erg (not shown in Fig.4, but see Fig.7). 

Combining adjacent pixels to events is a non-trivial matter. Synchrony between peaks is required and a tolerance must be specified. Tolerances of $\pm$1 and $\pm$3 minutes have been used in the literature (cf. Tab.1). At a pixel size of 1900 km, they correspond to lower limits of accepted motions of 16$N$ and 5.3$N$ km/s, where $N$ is the number of traversed pixels. $N$ may be up to about 7. Events exhibiting slower motions would be counted as several events. Thus in principle, a large synchrony tolerance is desirable to avoid breaking up big events into spurious fragments. We note however that only one moving event in the quiet corona has ever been described in detail (Benz \& Krucker 1998). Parnell \& Jupp (2000) and Berghmans et al. (1998) have noticed only few such events. Nearly all events emerge and decay on the same spot.

On the other hand, a large tolerance enhances the number of spurious associations. KB have reported an event rate of 0.0115 per pixel per minute above 3 $\sigma$. The number must be multiplied by the average event size, 1.63 pixels, to yield the chance association rate of an event with an other independent event. For the 4 closest pixels, the chance to have a coincident, but unrelated peak within a 2 minute time interval (tolerance of $\pm$1 minute) is 15\%. It becomes 3 times larger and is excessive for a tolerance of $\pm$3 minutes. 

\begin{figure} \plotone{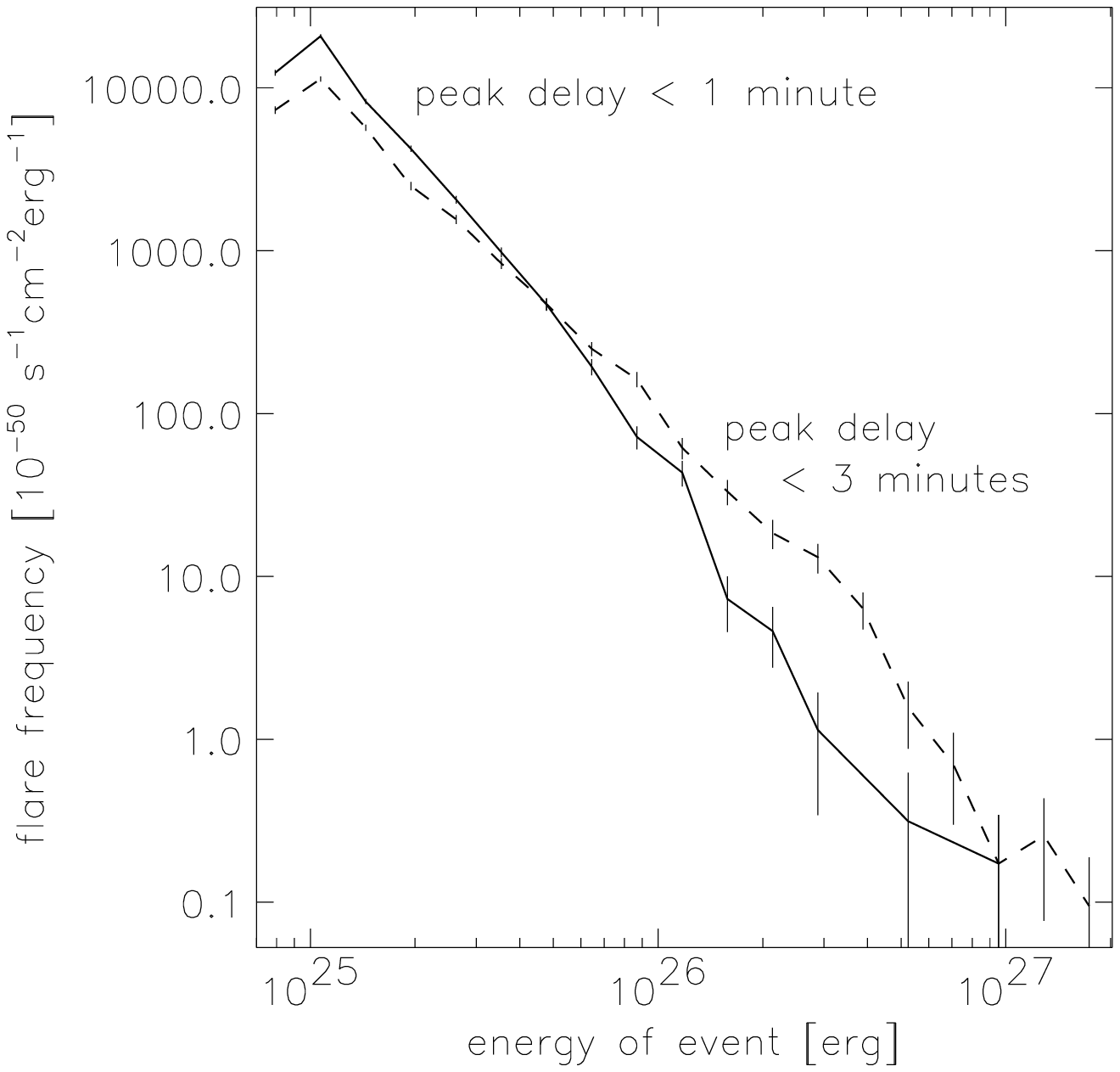}
\caption[f5.eps]{The thermal energy distribution of micro-events for two different tolerances in the timing of adjacent pixel's peak.  The power-law index decreases from $2.59\pm0.02$ for the $\pm$1 minute tolerance to $2.15\pm0.02$ for the $\pm$3 minute tolerance for combining events in adjacent pixels (height model: $s_{eff}=5000$km). \label{fig5}}
\end{figure}
 
The effect of increasing the tolerance from  $\pm$1 to $\pm$3 minutes is evident in Fig.5. The power-law index decreases by 0.44 because small events are combined to bigger ones. The effect is most pronounced at small events where the ratio of circumference to area is largest. The accumulation process is also documented by Fig.6. The largest events found using a large tolerance appear to be unrealistic agglomerations of nearly the size of supergranular cells.

\begin{figure}
\plotone{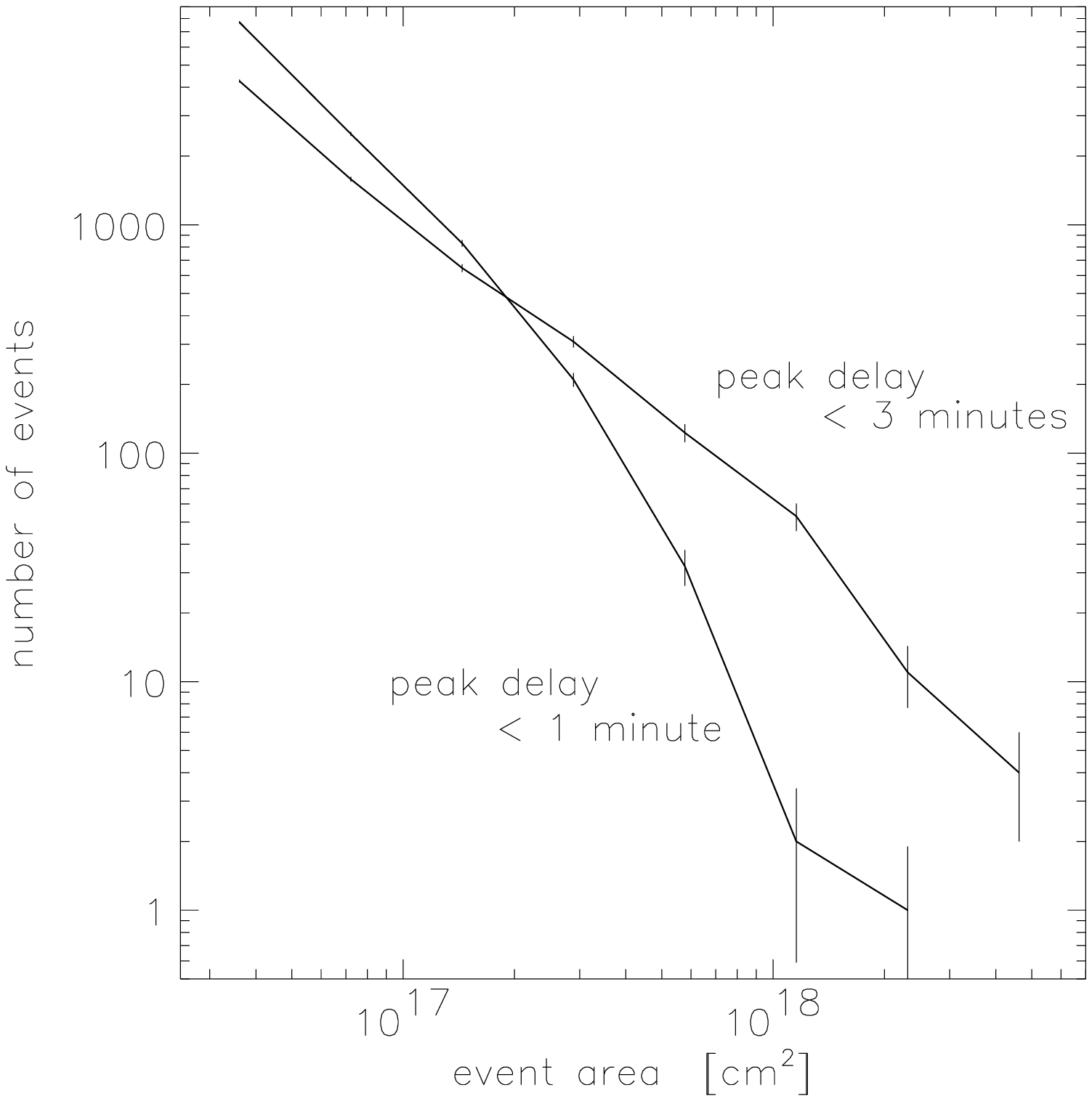}
\caption[f6.eps]{The area distribution of impulsive heating events observed in the quiet corona. The number of very large events strongly increases from the $\pm$1 minute requirement to the $\pm$3 minute requirement for the simultaneity of adjacent peaks.\label{fig6}}
\end{figure}

There is apparently no simple answer for the correct tolerance. More sophisticated conditions for combining pixels to events have to be developed based on the experience of well studied events. If  there is a tolerance interval that balances chance associations and break-up, we expect it to be at the lower end of the published range. Thus the power-law indices derived with a tolerance of $\pm$3 minutes and more should be corrected up.

\begin{figure}
\plotone{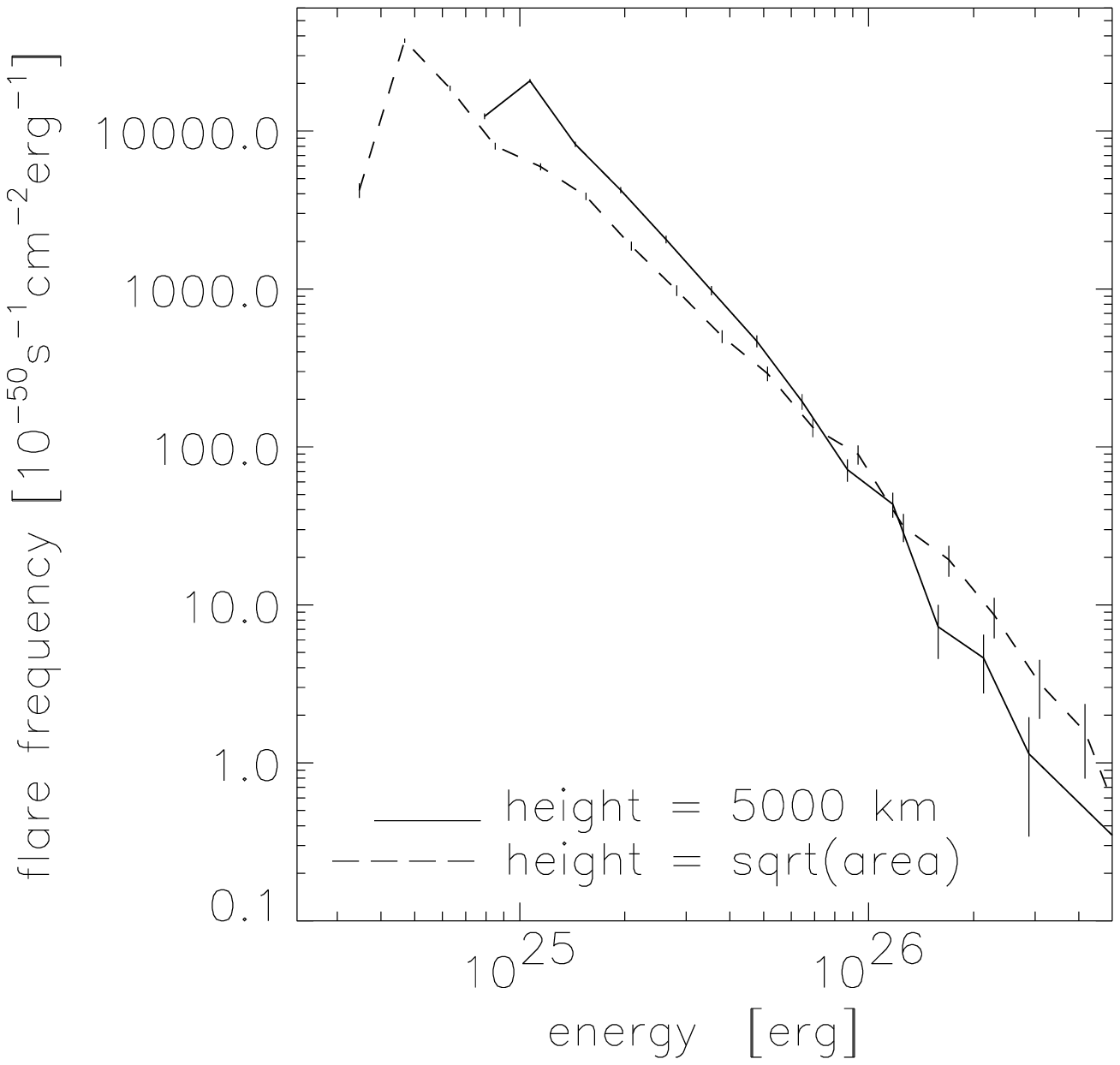}
\caption[f7.eps]{The frequency distribution of micro-event energy for different assumptions on the effective height. The model with a constant height (full) curve is compared with a model assuming the height given by the square root of the area. \label{fig7}}
\end{figure}

\subsection{Height Model}
The effective height $s_{eff}$ (Eq.6) has been modeled in the literature by a constant in energy and, alternatively, by $b\sqrt{A}$, where $b$ is some constant and $A$ the observed event area. Figure 7 shows the result for the same data evaluated with different height models. The power-law index decreases from 2.59 for a constant height to 2.31 for the $\sqrt{A}$-model in agreement with the theoretical derivations of Mitra Kraev \& Benz (2001, Eq.25). 

Constant $s_{eff}$ has been assumed in single pixel analyses and in KB. The $\sqrt{A}$-model is more plausible if events across a large energy range are compared and should be preferred. 

\begin{figure}
\plotone{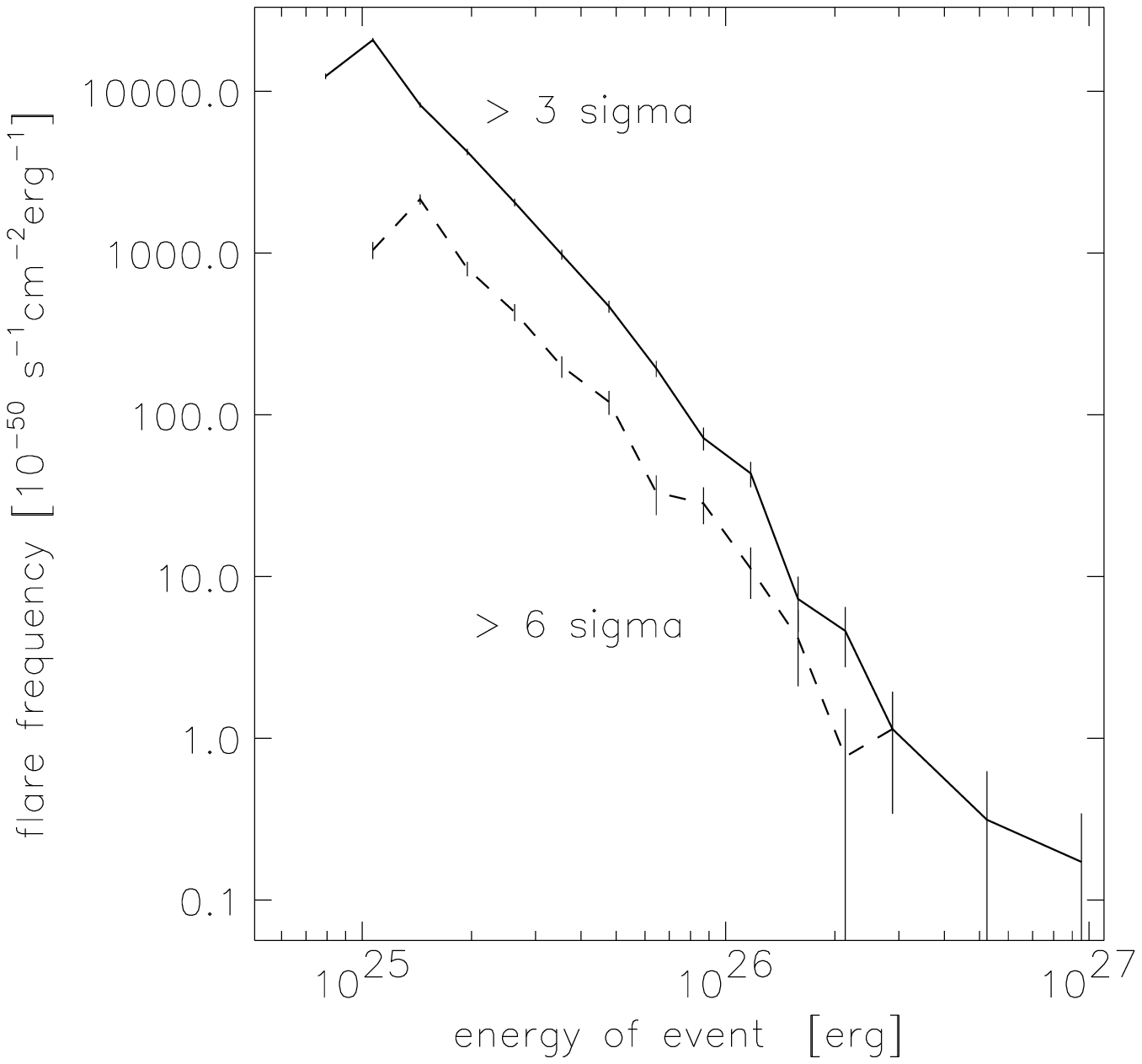}
\caption[f8.eps]{The energy distribution for different cutoffs in $\sigma$ for the detection and combination of peaks in adjacent pixels. The $>$6$\sigma$ curve (dashed) represents the results for an instrument with half the sensitivity of EIT.\label{fig8}}
\end{figure}

\begin{figure}
\plotone{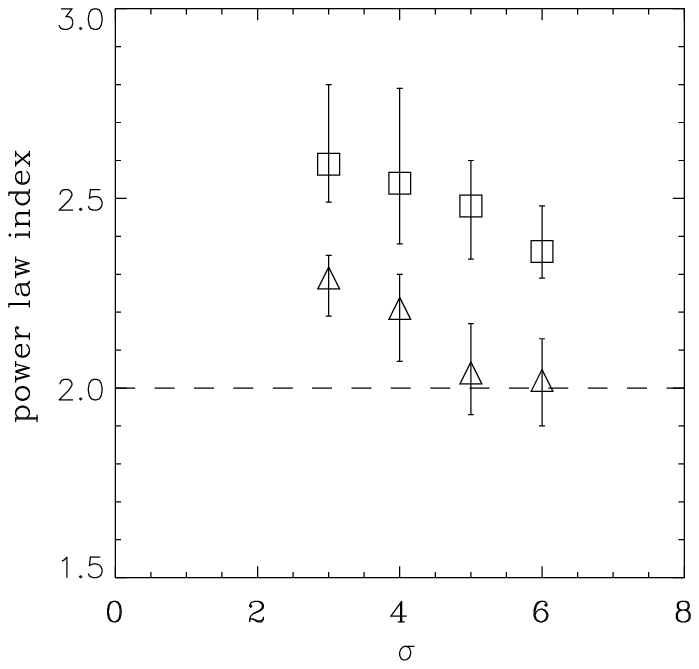}
\caption[f9.eps]{The power-law index of the energy distribution vs. different cutoffs in $\sigma$ for the detection and combination of peaks in adjacent pixels. The symbols denote the two different models used: $\Box$ for constant effective height of all events, $\bigtriangleup$ for height equal to the square root of the area. A tolerance in timing peaks of adjacent pixels of $\pm$1 minutes was used. The error bars are estimates of the total possible error.\label{fig9}}
\end{figure}

\subsection{Sensitivity}
The power-law index from EIT observations is 2.2 - 2.3 (Tab.2, $\sqrt{A}$-model), whereas for TRACE a range of 2.04 - 2.13 has been reported (Parnell \& Jupp 2000). Figure 8 suggests a possible interpretation. Apparently, the slope of the energy distribution also depends on the cutoff used to select peaks. The exponent changes from 2.59 to 2.39 if the $>3\sigma$ condition is enhanced to $>6\sigma$. The energy cutoff is shifted to higher energies, and, most notably, the number of micro-events at a given energy is lower by a factor of 4.2. The reason for this is that the higher $\sigma$ cutoff eliminates some of the pixels with low-level variation at the limb of events. Thus the area of an event tends to be reduced and so does the derived energy. The distribution in Fig. 8 is shifted to lower energy, and its cutoff to higher minimum energy. The effect is slightly more efficient at lower energies where events have a generally smaller signal-to-noise ratio.

The difference between EIT and TRACE results may be an effect of sensitivity. Figure 9 shows a decrease of $\delta$ with increasing cutoff in units of $\sigma$. In both cases the power-law index continuously decreases with $\sigma$. The $\sigma$ for EIT is given by photon noise and is 4.4 times smaller than for TRACE, where is it determined by cosmic ray hits and their cleaning (Parnell \& Jupp 2000; Aschwanden et al. 2000a). This agrees qualitatively with Fig. 9, but the effect is smaller than predicted, possibly reduced by the higher spatial resolution of TRACE.

Despite this general agreement on the power-law index, we note a strong discord with Parnell \& Jupp (2000) in the absolute value of the distribution at a given energy. It seems to be lower by more than an order of magnitude relative to KB. The distribution is even lower than in Aschwanden et al.(2000b), who have reduced their event number by severe event selections. There is no agreement yet for the discrepancy, but most likely it is a simple numerical error.

\begin{table}
\begin{center} 
\caption{Power-law indices of the energy distribution of micro-flares derived by various recent authors from different data and for different model assumptions, tolerance of synchrony in minutes, and flare selection. The reported range for power-law fits at different energies is given.} 
\begin{tabular}{lrrrr} \hline
& & & & \\
reference & $h$=const&$h=\sqrt{A}$&$h=\sqrt{A}$&$h=\sqrt{A}$ \\
  & $\Delta t =\pm$ 1 &$\Delta t =\pm$ 1 &$\Delta t =\pm$ 3& $\Delta t =\pm$ 3\\
& & & &selection\\
\hline
PJ (2000)&2.4 - 2.6&2.0 - 2.1& & \\
AEA (2000)&2.3 - 2.5&2.2 - 2.3&2.1 - 2.2&1.8 - 1.9 \\
KB (1998) + this work&2.3 - 2.6&2.2 - 2.4&2.0 - 2.2&  \\
& & & & \\
\hline
\end{tabular}                                             

\end{center}
\end{table}

\section{Discussion}
This second look at the EIT data has shown general agreement with TRACE, if the same methods are used (Tab.2). As a best approach, we suggest a small tolerance ($\Delta t = \pm 1$ min) in the timing of adjacent pixels to be summed up to events. A height model that assumes a smaller effective line-of-sight thickness for smaller events is probably more appropriate ($h=\sqrt{A}$). Then a power-law index of $\delta$ = 2.3$\pm 0.1$ results from EIT data (third column in Tab.2). 

The above height model reduces the observed thermal input energy by micro-flares from 16\% (KB) to about 5\% of the radiated output ($\gapprox 10^6$K). To account for the nanoflare heating scenario, the observed micro-events have to either continue to much lower energies or to provide more energy per event than the measured thermal part, or both.

The simple considerations on energy inputs and outputs in Sect. 2 suggest that this percentage may have to be multiplied by a factor of $\alpha(\beta +\gamma)/\mu \approx 3$ to include the other forms of energy involved in micro-events. We note that this factor has an upper limit given by the observed energy output. The more energy the observed micro-events contain, the less micro-events below the current energy threshold may contribute. The latter constitute a part of the unresolved background in the "nanoflare heating" scenario.  Thus the background between observable micro-events may eventually fall below the observed minimum value if the energy input were mostly by the larger events. The observed emission measure time profile of single pixels does not show such gaps. On the contrary, as illustrated in Fig.1 the discernible events reside on top of a background. In numerical simulations assuming $\alpha/\mu$=1, Mitra Kraev \& Benz (2001) find a best fit of the observed Fourier power spectrum in time with $\beta +\gamma \approx 1.2$ (adapted to the average temperature found here). 

The factor $\alpha > 1$ is different from the others as it constitutes a non-impulsive, possibly non-local heating enhancing the background level. A scenario is conceivable where waves travel from the energy release site across magnetic field lines for tens of pixels ($\gapprox 10^4$km). The superposition of waves from many events could form a quasi-steady turbulent background. Such an energy input amounts to "dark power" as it does not manifest itself in variability.  Its dissipation would heat the coronal background emission measure without changing the power spectrum. The low-energy cutoff of the energy distribution would then have to be accordingly higher to fit the observed level. 

In view of the simulations and the estimate of $\alpha \approx 2$, the micro-events observed in the thermal energy range $5\times 10^{24} - 5\times 10^{26}$erg (Fig.7) suggest a total power input of about 12\% relative to the inferred output of the same quiet-sun field of view.

\section{Conclusions}
Nanoflares have originally been introduced as simple energy inputs retaining the high temperature of the corona. Recent observations indicate persistently that micro-events behave much more like real flares, heating initially cold material and increasing the material content of the low corona. As shown by Brown et al. (2000), the transition layer reacts very sensitively to coronal heating and particle impact. Most easily observable in coronal heating is not the primary energy discharge that may well occur in the corona, but the reaction of the chromosphere and the evaporation of hot material. The latter has apparently been observed in micro-events of the quiet sun. Nevertheless, the fact that they are secondary phenomena must not be forgotten when the energy of the whole event is estimated. 

In particular the expansion energy, losses during the rise time and the release of other forms of energy have to be added. Micro-events are not wave-like, but waves may be excited by the events that travel across field lines and into the upper corona. Although the thermal energy of observed micro-events is not sufficient for the coronal heating, micro-events are now the best source of information and evidence on the heating process of the quiet corona.

Individual micro-events have many physical properties in common with regular flares. However, their location and statistical behavior are different. The observed variability of the quiet corona takes place at low altitudes, mostly below 10$^4$ km. This is different from active regions, where extremely bright, quasi-stationary loops extend to more than an order of magnitude larger altitude. The quiet corona at low height appears simpler and more uniform from the point of view of loop length and temperature range. In any case, the micro-events are a less sporadic phenomenon than regular flares and are not known to correlate with solar activity. Therefore, contrary to the suggestions by Aschwanden et al. (2000b, Fig.10), the energy distribution of micro-events cannot be directly compared with regular flares. Micro-events in quiet regions and flares in active regions form statistically different populations.

It may be noted here that the variability even increases at lower temperatures (Berghmans et al. 1998) and has a maximum in the transition region, where the time scales are shorter and the temporal power spectrum is flattest (Benz \& Krucker 1999). This suggests that a more complete modeling of flare-like events at the interface of corona and chromosphere is highly desirable. As the observed variability is not wave-like, it is the best indicator for the relevance of the "nanoflare scenario". Mitra Kraev \& Benz (2001) have proven by simulations the existence of a nanoflare model that can explain all observed facts relevant to the heating of the quiet corona. Nanoflare heating thus cannot be excluded as some previous authors have claimed. 

The time of quick progress in micro-events is clearly over. Low-corona variability has opened a new, vast and largely still unexplored field. Only studies that investigate carefully the persisting problems can bring progress. One of them will be the detailed investigation of even smaller single events at high spatial and temporal resolution.

\acknowledgments
We thank C.E. Parnell and D. Berghmans for a critical reading of the manuscript and M.J. Aschwanden for clarifying discussions. The work has profited from the participation at ISSI workshops in Bern on Subresolution Solar Physics (1999/2000) and used data from SoHO/EIT, funded by CNES, NASA, and the Belgian SPPS. The work at ETH Zurich is financially supported by the Swiss National Science Foundation (grant No. 2000-061559). SK is supported by the Swiss National Science Foundation (grant No. 8220-056558) and by NASA grants NAG5-2815 and NAG5-6928 at Berkeley.


\clearpage
\begin{thebibliography}{}
\bibitem{} Aschwanden, M.J., et al. 2000a, ApJ, 535, 1027
\bibitem{} Aschwanden, M.J., et al. 2000b, ApJ, 535, 1047
\bibitem{} Benz, A.O., \& Krucker, S. 1998, Sol. Phys., 182, 349
\bibitem{} Benz, A.O., \& Krucker, S. 1999, A\&A 341, 286
\bibitem{} Berghmans, D., Clette, F., \& Moses, D. 1998, A\&A, 336, 1039
\bibitem{} Brkovi\'c, A., et al. 2000, A\& A 352, 1083
\bibitem{} Brown, J. C., Krucker, S., G\"udel, M., \& Benz, A.O., 2000, A\&A, 359, 1185
\bibitem{} Brueckner, G.E., \& Bartoe, J.-D.F. 1983, ApJ,  272, 329
\bibitem{} Delaboudini\`ere, J.-P., et al. 1995, Sol. Phys., 162, 291 
\bibitem{} Dere, K.P., 1994, Adv. Space Res., 14, 13
\bibitem{} Gary, D.E., Hartl, D.M., \& Shimizu, T. 1997, ApJ, 477, 958
\bibitem{} Golub, L., Krieger, A.S., Vaiana, G.S., Silk, J.K., \& Timothy, A.F. 1974, ApJ, 437, 522
\bibitem{} Habbal, S.R. 1992, Ann. Geophys., 10, 34
\bibitem{} Handy, B., et al. 1999, Sol. Phys., 187, 229
\bibitem{} Harrison, R.A. 1997, Sol. Phys., 175, 467
\bibitem{} Koutchmy, S., Hara, H., Suematsu, Y., \& Reardon, K. 1997, A\& A, 320, L33
\bibitem{} Krucker, S., Benz, A.O., Acton, L.W., \& Bastian, T.S. 1997, ApJ, 488, 499
\bibitem{} Krucker, S., \& Benz, A.O. 1998, ApJ, 501, L213 (BK)
\bibitem{} Krucker, S., \& Benz, A.O. 2000, Sol. Phys. 191, 341
\bibitem{} Lin, R.P., Schwartz R.A., Kane S.R., Pelling R.M., \& Hurley K.C. 1984, ApJ 283, 421
\bibitem{} Mitra Kraev, U., \& Benz, A.O. 2001, A\& A 373, 318
\bibitem{} Parker, E.N. 1983, ApJ, 264, 642
\bibitem{} Parnell, C.E., \& Jupp, P.E. 2000, ApJ 529, 554
\bibitem{} Porter, J.G., et al. 1987, ApJ 323, 380
\bibitem{} Pre\'s, P., \& Phillips, K.J.H. 1999, ApJ, 510, L73
\bibitem{} Pre\'s, P., Phillips, K.J.H., \& Falewicz, R. 2001, ApJ, in preparation
\bibitem{} Priest, E.R. \& Forbes, T. 2000, {\it Magnetic Reconnection}, Cambridge Univ. Press
\bibitem{} Rutten,R.J., \& Hogenaar, H.J. 2001, {\it Encyclopedia of Astronomy and Astrophysics}, IOP Bristol, p. 332
\bibitem{} Ulmschneider, P., Rosner, R., \&  Priest, E.R. (eds.), 1991, 
{\it Mechanisms of chromospheric and coronal heating}, Springer-Verlag, Berlin
\bibitem{} Winebarger, A.R., Emslie, A.G., Mariska, J.T., \& Warren H.P. 2001, ApJ, submitted
\end{thebibliography}
\end{document}